\documentclass[superscriptaddress,aps,journal=prl,preprint]{revtex4-2}
\usepackage{graphicx}
\usepackage{epstopdf}
\usepackage{bm,amsmath,amssymb} 
\usepackage{color, soul} 
\usepackage{dcolumn}   
\usepackage{multirow}
\usepackage{pifont}
\usepackage[colorlinks=true, linkcolor=black, citecolor=black, urlcolor=blue]{hyperref}
\newcommand{\etal}{{\em et al}.\ }
\newcommand{\eg}{{\em e.g.}}

\newcommand{\POP}{$\mathcal{P}_{\rm OP}$}
\newcommand{\PIP}{$\mathcal{P}_{\rm IP}$}
\newcommand{\EOP}{$\mathcal{E}_{\rm OP}$}
\newcommand{\EIP}{$\mathcal{E}_{\rm IP}$}
\newcommand{\ainse}{$\alpha$-In$_2$Se$_3$}
\newcommand{\binse}{$\beta$-In$_2$Se$_3$}
\newcommand{\dwas}{DW$_{A,s}$}
\newcommand{\dwal}{DW$_{A,l}$}

\usepackage{lineno}
\begin{document}
\title{Intrinsic ferroelectric switching in two-dimension $\alpha$-In$_2$Se$_3$}
\author{Liyi Bai}
\thanks{These three authors contributed equally}
\affiliation{Key Laboratory for Quantum Materials of Zhejiang Province, Department of Physics, School of Science, Westlake University, Hangzhou, Zhejiang 310024, China}
\affiliation{Institute of Natural Sciences, Westlake Institute for Advanced Study, Hangzhou, Zhejiang 310024, China}
\author{Changming Ke}
\thanks{These three authors contributed equally}
\affiliation{Key Laboratory for Quantum Materials of Zhejiang Province, Department of Physics, School of Science, Westlake University, Hangzhou, Zhejiang 310024, China}
\affiliation{Institute of Natural Sciences, Westlake Institute for Advanced Study, Hangzhou, Zhejiang 310024, China}
\author{Zhongshen Luo}
\thanks{These three authors contributed equally}
\affiliation{School of Physical Science and Technology, Jiangsu Key Laboratory of Thin Films, Soochow University, Suzhou, 215006, China}
\author{Tianyuan Zhu}
\affiliation{Key Laboratory for Quantum Materials of Zhejiang Province, Department of Physics, School of Science, Westlake University, Hangzhou, Zhejiang 310024, China}
\affiliation{Institute of Natural Sciences, Westlake Institute for Advanced Study, Hangzhou, Zhejiang 310024, China}
\author{Lu You}
\email{lyou@suda.edu.cn}
\affiliation{School of Physical Science and Technology, Jiangsu Key Laboratory of Thin Films, Soochow University, Suzhou, 215006, China}
\author{Shi Liu}
\email{liushi@westlake.edu.cn}
\affiliation{Key Laboratory for Quantum Materials of Zhejiang Province, Department of Physics, School of Science, Westlake University, Hangzhou, Zhejiang 310024, China}
\affiliation{Institute of Natural Sciences, Westlake Institute for Advanced Study, Hangzhou, Zhejiang 310024, China}

\date{\today}

\begin{abstract}
Two-dimensional (2D) ferroelectric semiconductors present opportunities for integrating ferroelectrics into high-density ultrathin nanoelectronics. Among the few synthesized 2D ferroelectrics, \ainse, known for its electrically addressable vertical polarization has attracted significant interest. However, the understanding of many fundamental characteristics of this material, such as the existence of spontaneous in-plane polarization and switching mechanisms, remains controversial, marked by conflicting experimental and theoretical results. Here, our combined experimental characterizations with piezoresponse force microscope and symmetry analysis conclusively dismiss previous claims of in-plane ferroelectricity in \ainse.~The processes of vertical polarization switching in monolayer \ainse~are explored with deep-learning-assisted large-scale molecular dynamics simulations, revealing atomistic mechanisms fundamentally different from those of bulk ferroelectrics. 
Despite lacking in-plane effective polarization, 1D domain walls can be moved by both out-of-plane and in-plane fields, exhibiting unusual avalanche dynamics characterized by abrupt, intermittent moving patterns. The propagating velocity at various temperatures, field orientations, and strengths can be statistically described with a universal creep equation, featuring a dynamical exponent of 2 that is distinct from all known values for elastic interfaces moving in disordered media. 
This work rectifies a long-held misunderstanding regarding the in-plane ferroelectricity of \ainse, and the quantitative characterizations of domain wall velocity will hold broad implications for both the
fundamental understanding and technological applications of 2D ferroelectrics.
\end{abstract}

\maketitle
\newpage

\section{Introduction}
For a ferroelectric thin film, the depolarization field that arises due to the imperfect screening of polarization bound charges on surfaces is inversely proportional to the thickness of the film~\cite{Junquera03p506}. The suppressed polarization along the direction of reduced dimensionality has been a main obstacle for the miniaturization of ferroelectric-based devices. Among the limited number of synthesized two-dimensional (2D) ferroelectrics,  $\alpha$-In$_2$Se$_3$ has garnered considerable attention, mainly because of its advantageous out-of-plane polarization (\POP) in the monolayer limit~\cite{Zhou17p5508,Cui18p1253,Xiao18p227601}. This characteristic allows for full utilization of the atomic thickness, thereby opening avenues for the development of high-performance, ultrathin nanoelectronics~\cite{Wan19p1808606}. Notably, the 2D ferroelectricity in $\alpha$-In$_2$Se$_3$ was first predicted theoretically based on density functional theory (DFT) calculations~\cite{Zhou17p5508}, which was later confirmed experimentally~\cite{Cui18p1253,Xiao18p227601}. 
Several device prototypes, including ferroelectric channel transistor~\cite{Wan22p25693} and synaptic ferroelectric semiconductor junction~\cite{Si19p580}, have been fabricated using quasi-2D In$_2$Se$_3$ films with a thickness of tens of nanometers. In addition to forming the $\alpha$ phase, In$_2$Se$_3$ can crystallize into the $\beta'$ phase with in-plane polarization ($P_{\rm IP}$) and the paraelectric $\beta$ phase (Fig.~\ref{Fig1}\textbf{a})~\cite{zheng18peaar7720,Zhang19p8004,Zheng22peabo0773,Chen23p1077}. Leveraging both polymorphism and ferroelectricity of In$_2$Se$_3$ establishes a versatile platform for unlocking new functionalities~\cite{Han22p55}.

Recent experiments have demonstrated that van der Waals (vdW) stacked bilayers of nonferroelectric monolayers can be engineered into ferroelectrics through sliding and twisting~\cite{ViznerStern21p1462,Roge22p973,Tsymbal21p6549,Wu21pe2115703118,Wang23p542}, thus substantially expanding the family of 2D ferroelectrics possessing \POP.
Much like their bulk counterparts, the functional attributes of 2D ferroelectrics depend critically on the polarization response to external stimuli, whereas it is not clear whether various switching models developed for bulk ferroelectrics can be directly applied to reduced dimensions~\cite{Zhang22p25}. 
At the most fundamental level, the controlled engineering of ferroelectric switching in 2D hinges on a quantitative characterization of the dynamics of 1D domain walls, which are nanoscale interfaces between differently polarized 2D domains. This endeavor has been particularly challenging due to the high temporal and spatial resolution required for both experimental and theoretical approaches, such that there is virtually no reported data on the velocity of 1D domain walls in 2D ferroelectrics. In this work, we use 2D $\alpha$-In$_2$Se$_3$ as a model system to investigate the switching mechanisms in 2D.

The structural origin of \POP~in  monolayer \ainse~(space group $P3m1$) is unambiguous: it results from the shift of the central Se layer along the $z$-axis that breaks the out-of-plane inversion symmetry (Fig.~\ref{Fig1}\textbf{b}). Despite extensive studies, several specifics of ferroelectricity in 2D  \ainse~remain unclear. Particularly, there is an on-going debate regarding whether monolayer \ainse~exhibits electrically switchable in-plane effective polarization (\PIP). 
In the original paper predicting 2D ferroelectricity in monolayer \ainse~\cite{Ding17p14956}, it was argued that it should also possess \PIP, mainly deduced from the cross-sectional crystal structure which suggests a lateral off-centred displacement of Se atom.
However, as pointed out by follow-up studies~\cite{Xiao18p1707383}, the C$_{3\rm v}$ point group symmetry (out-of-plane three-fold rotational axis) of \ainse~strictly prohibits in-plane switchable polarization (also refereed to as {\em effective polarization} in modern theory of polarization~\cite{KingSmith93p1651, Resta93p1010}), according to Neumann's principle. This is evident from the top view of \ainse, similar to $h$-BN (Fig.~S1), which is invariant with respect to a three-fold rotation thus nullifying any in-plane spontaneous polarization.  Nevertheless, multiple experimental studies have reported switching-like behaviors when in-plane electric fields (\EIP) are applied to \ainse~thin films~\cite{Xue19p1901300,Wang20p45}.

Another experiment-theory conundrum for 2D \ainse~involves the coercive field that switches \POP. Previous DFT investigations have indicated a rather low switching barrier of $\approx$66~meV per unit cell (uc) in monolayer \ainse~\cite{Ding17p14956}, yet the observed out-of-plane switching field is commonly above 1 V/nm (10 MV/cm)~\cite{Xiao18p227601}, several orders of magnitude higher than that for PbTiO$_3$-based perovskite ferroelectrics  ($\approx$10 kV/cm)~\cite{Lente01p5093} which actually have larger switching barriers ($\approx$200~meV/uc). The low barrier predicted by DFT does not support  ``dipole locking mechanism", which proposes that the locked out-of-plane and in-plane motion of the middle Se atom would lead to a large barrier~\cite{Xiao18p227601}. We note that this terminology has its own issues, as it could be mistakenly interpreted as the locking
between out-of-plane and in-plane net dipoles, despite the original work explicitly stating that there is only in-plane asymmetry (see additional discussions in Supplementary Sect.~I). This asymmetry does not lead to in-plane net electric dipole (due to three-fold rotational symmetry), but it allows for effective in-plane second-order optical dipole emission.
Such experiment-theory conundrum appears to be a common feature for vdW bilayers exhibiting sliding ferroelectricity. For example, the DFT-predicted switching barrier for Bernal-stacked $h$-BN possessing sliding ferroelectricity is only 8 meV/uc~\cite{Li17p6382}, yet the experimental switching field is on the order of 0.1 V/nm~\cite{ViznerStern21p1462, Yasuda21p1458}.

To summarize, currently there is no consensus, be it experimental or theoretical, on several key questions:
(i) Does monolayer \ainse~exhibit switchable \PIP?
(ii) Why does 2D \ainse, similar to vdW bilayers of stacking-engineered ferroelectricity, exhibit such a high out-of-plane switching field? (iii) What are the polarization switching dynamics involved in 2D domains and 1D walls? Here, we address these questions with both experimental characterizations and a multiscale modeling approach that combines symmetry analysis, DFT calculations, and deep-learning-assisted large-scale molecular dynamics (MD) simulations.


\section{Results}
\subsection{Symmetry analysis of \ainse}
Monolayer \ainse~is in the space group of $P3m1$, while the bulk forms adopting 3R and 2H stacking orders are in the space groups of $P6_3mc$ and $R3m$, respectively. All these space groups exhibit a three-fold rotational axis along $z$ (the out-of-plane direction for the monolayer). 
Proper symmetry analysis should lead to a straightforward conclusion: there is only
in-plane piezoelectricity due to the in-plane inversion symmetry breaking, but there
is no in-plane ferroelectricity. The in-plane asymmetry of \ainse~is similar to that in 2H or 3R MoS$_2$ and $h$-BN.
It is possible that residual uniaxial strains in \ainse~samples could disrupt the three-fold rotational symmetry, and the piezoelectric effect gives rise to finite but non-reversible \PIP.

The resilience of \POP~against the depolarization field in 2D \ainse~can also be understood with a symmetry argument, viewing the switching process from the ferroelectric to the reference paraelectric phase as an electric field-driven phase transition.
The Landau theory of group-subgroup structural phase transitions presumes that transitions result from the condensation of either a single or a group of collective degrees of freedom, conforming to a single irreducible representation (irrep) of the space group for the high-symmetry phase.
Within a space group, atoms occupy specific positions known as Wyckoff positions (or orbits), defined by their symmetry properties within the space group. Wyckoff orbit splitting refers to the phenomenon where a set of equivalent atomic positions in a crystal becomes distinct and non-equivalent due to a reduction in symmetry during a group-subgroup phase transition. Importantly, each type of group-subgroup transition has its own set of restrictions on how these splittings can occur (see the example of the paraelectric-ferroelectric phase transition of PbTiO$_3$ in Fig.~S2).
We find that though paraelectric \binse~(space group $P\bar{3}m1$) and ferroelectric \ainse~(space group $P3m1$) conform to the group-subgroup relationship, the $\beta \leftrightarrow \alpha$ transition cannot be realized by a structural distortion associated with an irrep of $P\bar{3}m1$, because the atomic occupations in \binse~and \ainse~do not conform to the required splitting of Wyckoff positions. Specifically, the In atoms occupy the 2d Wyckoff orbits in \binse, while the two In atoms in \ainse~occupy 1a and 1c orbit, respectively, violating the allowed splitting rule of 2d$\rightarrow$1b/1c associated with $P\bar{3}m1\rightarrow P3m1$ (see Fig.~\ref{Fig1}\textbf{a}-\textbf{b}). 
This is fundamentally different from (proper) perovskite ferroelectrics, where a polar soft mode identified by an irrep serves as the primary order parameter responsible for the paraelectric-ferroelectric phase transition~\cite{Cochran59p412}. 
Given that a depolarization field mainly suppresses the displacive distortion associated with the polar soft mode, we propose that \POP~in 2D \ainse~is protected by a principle we call ``splitting restriction", which refers to the forbidden splitting of Wyckoff orbits during the postulated phase transition. We highlight that the same mechanism also plays a crucial role in the stabilization of \POP~emerged in sliding and moir\'e ferroelectricity~\cite{Lemerle98p849, Paruch05p197601, Paruch13p667,Ji23p146801} (see detailed discussions about Bernal-stacked $h$-BN in Supplementary Sect.~VI). 
Additionally, the splitting restriction principle predicts that $\beta\rightarrow\beta'$ could be favored over $\beta\rightarrow\alpha$, which is consistent with our Raman spectroscopy characterizations of 2H-\ainse~during a heating-cooling cycle (see Supplementary Sect.~II).

\subsection{Piezoelectric response microscopy characterizations of \ainse}
The aforementioned symmetry analysis resoundingly indicates the absence of \PIP~in \ainse, contrasting to a flurry of research studies that have claimed its existence.
Nearly all reports of the presence of \PIP~in 2D \ainse~depended on measurements of piezoresponse force microscopy (PFM) with the Dual AC Resonance Tracking (DART) mode\cite{Cui18p1253,Yuan19p1775,Xue18p4976,Xue19p1901300,Li20p2000061,Wang20p45}.
We will first argue, based on fundamental PFM theory, that acquiring in-plane PFM (IP-PFM) signal in \ainse~is not possible.
In IP-PFM studies, the piezoresponse signal is related to the torsional motion of the atomic force microscopy (AFM) probe, which is caused by the shear strain of the sample under vertical electric field. The relevant piezoelectric coefficient in this process is $d_{35}$ (or $d_{34}$), where the subscript 5 (or 4) denotes the induced shear strain in $xz$ (or $yz$) plane under Voigt notation (Fig.~S4). However, for monolayer \ainse with a space group $P3m1$, both $d_{35}$ and $d_{34}$ components are zero within its piezoelectric tensorial matrix. The same situation applies to  bulk \ainse~with 2H or 3R stacking order (Fig.~S5).

\begin{figure}[t]
\centering
\includegraphics[width=6.4 in]{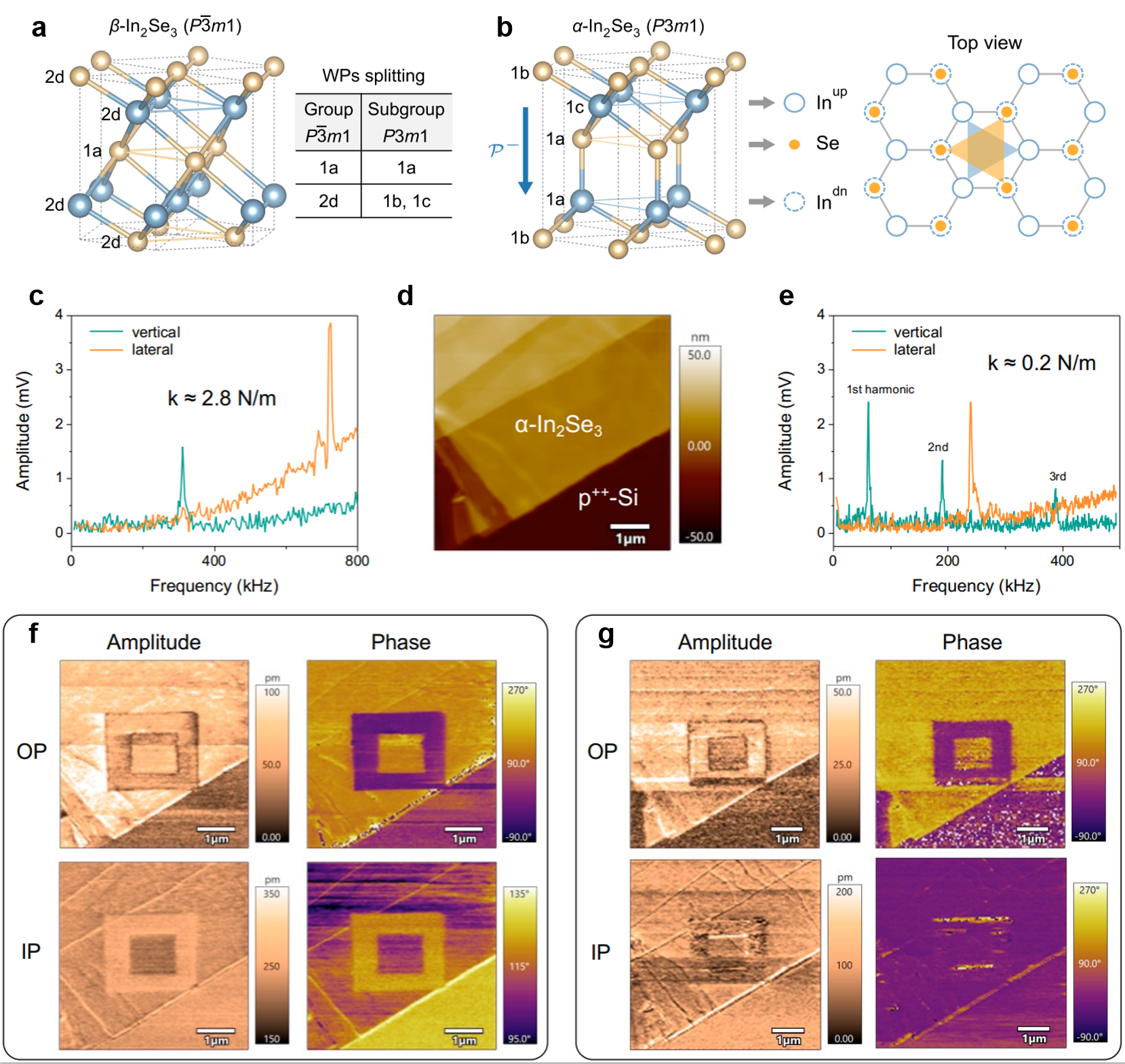}
\caption{{\bf OP- and IP-PFM measurements on 2H $\alpha$-In$_2$Se$_3$ using AFM probes with different force constants.} Schematics of \textbf{a,} paraelectric $\beta$ phase and \textbf{b,} ferroelectric $\alpha$ phase with Wyckoff positions (WPs) labeled.  For simplicity, the top view of $\alpha$-In$_2$Se$_3$ does not display the outermost Se atoms. 
The table presents the splitting rules of WPs for the phase transition of $P\bar{3}m1\rightarrow P3m1$. The In atoms in \ainse~occupy 1a and 1c orbit, respectively, which does not confirm to the 2d$\rightarrow$1b/1c splitting rule for the symmetry break $P\bar{3}m1\rightarrow P3m1$. \textbf{c,}  Vertical and lateral surface tuning spectra using a probe with \textbf{c,} $k\approx2.8$ N/m and \textbf{e,} $k\approx0.2$ N/m. \textbf{d,} AFM topographic image of the measured 2H $\alpha$-In$_2$Se$_3$ flake. \textbf{f,} OP- and IP-PFM amplitude and phase images acquired by the probe with \textbf{f,} $k  \approx 2.8$ N/m and \textbf{g,} $k  \approx 0.2$ N/m. 
 }
\label{Fig1}
\end{figure}

Second, it is well known in the PFM community that out-of-plane  and in-plane deformation signals are strongly convoluted in practical measurements~\cite{Piner02p3392,Hoffmann07p016101}. Thus, special care and a proper measurement protocol are required to unambiguously separate the vertical and lateral contributions with correct interpretations of the results, which were seriously lacking in previous studies of  \ainse~(see case-by-case analysis of the previous reports in Supplementary Sect.~III). Herein, we perform a comparative PFM study by using two AFM probes with different force constants. One probe has a force constant ($k$) of $\approx 2.8$ N/m, similar to those widely used in previous reports, while the other is much softer ($k\approx$0.2 N/m) with better torsional sensitivity~\cite{You18p1803249}. Both the out-of-plane and in-plane surface tuning spectra for both probes were recorded first to determine the contact resonance between the probe and the sample for the subsequent DART PFM measurements. As shown in Fig.~\ref{Fig1}\textbf{c} and \ref{Fig1}\textbf{e}, the resonant peaks for vertical deflection and lateral torsion modes are distinct from each other for both probes. 

Next, OP- and IP-PFM images are aquired using both probes driven by an AC voltage at their respective first harmonic frequencies. A box-in-box pattern was written by a DC biased probe in advance, as a reference for the following measurements. For OP-PFM, both probes provide strong piezoresponse amplitude on the \ainse~ surface compared with the Si substrate. Additionally, the phase images show sharp 180$^\circ$ difference between upward and downward domains, confirming the noticeable piezoresponse ($d_{33}$) in the out-of-plane direction. For IP-PFM, however, although we observe sizable ``amplitude signal”, it is not distinguishable between \ainse~and the non-piezoelectric Si substrate. As mentioned, this spurious signal originates from the electronic background of the lock-in amplifier at high frequency. The phase image obtained by the stiff probe exhibits a domain pattern identical to the OP-PFM image (Fig.~\ref{Fig1}\textbf{f}), yet the phase difference between oppositely polarized domains is much less than 180$^\circ$. Thus, the observed IP piezoresponse signal is probably a crosstalk effect from the OP signal. By using a soft probe, this effect is reduced, but still visible (Fig.~\ref{Fig1}\textbf{g}).

\begin{figure}[t]
\centering
\includegraphics[width=6.4 in]{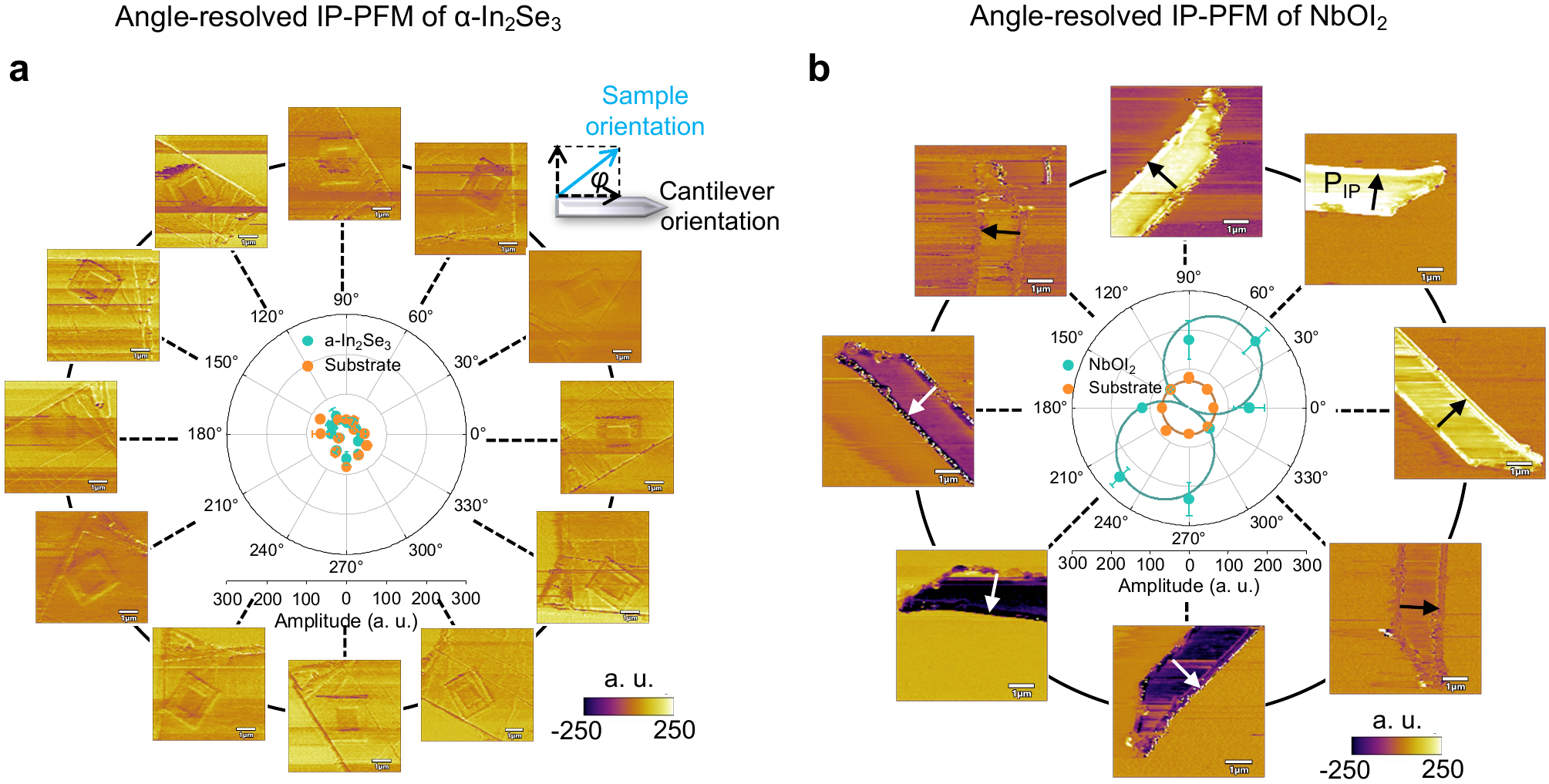}
\caption{{\bf A comparative study by angle-resolved IP-PFM.} Angle-resolved IP-PFM images of \textbf{a,} 2H $\alpha$-In$_2$Se$_3$ and \textbf{b,} NbOI$_2$. The polar plots in the center of \textbf{a} and \textbf{b} are the extracted mean amplitude of each PFM image at respective azimuthal angle. The measurement conditions were fixed for two samples. }
\label{Fig2-exp}
\end{figure}

Another important practice in PFM to confirm the intrinsic in-plane response is to perform an angular-resolved lateral PFM by rotating the sample at different azimuthal angles with regard to the AFM cantilever~\cite{You18p1803249,You12p5388}. 
We carry out a comparative study by conducting angle-resolved IP-PFM on both 2H \ainse~and NbOI$_2$ with known in-plane ferroelectricity under the same measurement condition, with the results summarized in Fig.~\ref{Fig2-exp}.  The measured IP-PFM amplitude is maximized when the direction of \PIP~is perpendicular to the cantilever axis and minimized when they are parallel to each other. Hence, we should expect the IP-PFM amplitude to follow an absolute sinusoidal function with a period of 180$^\circ$. This is exactly what is observed in the central polar plot of Fig.~\ref{Fig2-exp}\textbf{b} for NbOI$_2$, and the in-phase signal also changes its sign (phase reversal) when the sample is rotated by 180$^\circ$. In stark contrast, the 2H \ainse~sample with a pre-written box-in-box pattern shows no periodic modulation in both IP-PFM amplitude and phase (Fig.~\ref{Fig2-exp}\textbf{a}), suggesting the measured lateral response does not originate from intrinsic in-plane piezoresponse. Given the above symmetry analysis and experimental results, we can safely exclude the existence of \PIP~in 2H \ainse. 

The out-of-plane piezoresponse and polarization, on the contrary, is confirmed (Fig.~\ref{Fig1}) in our PFM investigations. We note that the out-of-plane switching field is extremely large, consistent with the prediction of splitting restriction principle. In an upward domain of a 30-nm-thick flake on highly-doped Si, it is only possible to obtain a minor hysteresis loop (incomplete switching) under a DC voltage up to 9 V (Fig.~S7\textbf{d}), while no sign of switching is observed in the downward domain (Fig.~S7\textbf{e}). At such a high field, irreversible surface modification due to the electrochemical process starts to take place (Fig.~S7\textbf{a}). The results indicate that the out-of-plane coercive field is possibly beyond 3 MV/cm. For comparison, vdW ferroelectric CuInP$_2$S$_6$ with a similar thickness shows a coercive field below 1 MV/cm under the same measurement condition (Fig.~S7\textbf{f}). Interestingly, we find that the switching field is greatly reduced when the biased AFM probe is scanning on the sample surface, which could explain the widely reported domain switching behaviors in the literature. As detailed by our theoretical calculations below, it is indeed much easier to drive polarization switching through domain wall motions with an in-plane electric field despite the lack of in-plane ferroelectricity in monolayer \ainse.

\subsection{Domain switching in 2D \ainse}
\begin{figure}[t]
\centering
\includegraphics[width=6.4 in]{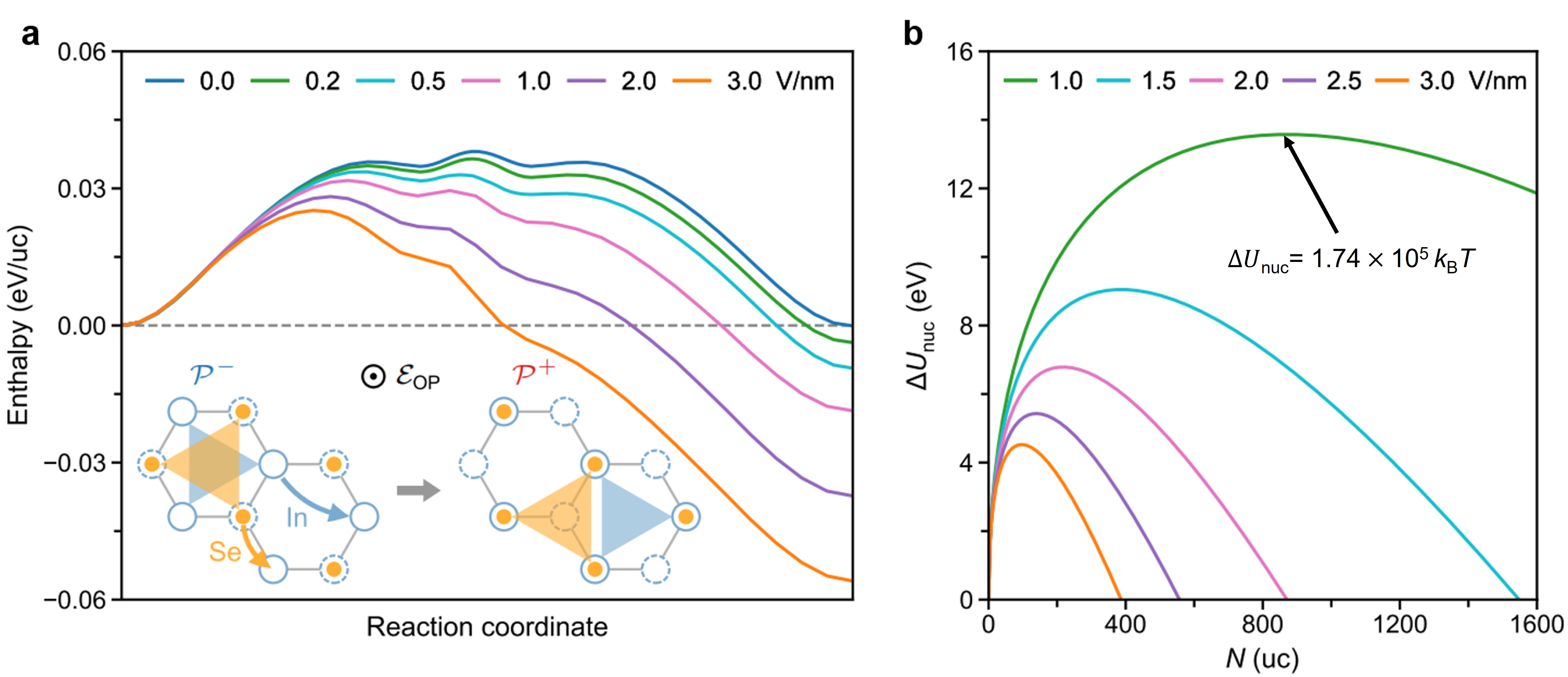}
\caption{{\bf Domain switching in monolayer \ainse.} \textbf{a,} Minimum energy paths of polarization reversal
identified with DFT-based NEB for various \EOP~magnitudes. The inset illustrates primary atomic movement patterns during this low-barrier switching pathway: entire layers of In and Se atoms need to move laterally in the same direction. This pathway is the same as the one reported in~\cite{Ding17p14956}. \textbf{b,} Analytical nucleation energy ($\Delta U_{\rm nuc}$) computed with DFT and MD parameters as a function of nucleus size ($N$) across various strengths of \EOP.}
\label{Fig3-theory}
\end{figure}

We now address the second experiment-theory conundrum: 
why experimentally a large out-of-plane electric field (\EOP) is required to switch \ainse, despite the identification of a low-barrier pathway using the DFT-based nudged elastic band (NEB) technique.
By carefully designing a pathway that connects the upward polarized and downward polarized monolayer \ainse, 
Ding~\etal~obtained a NEB barrier ($\Delta U^{\rm NEB}$) of 66~meV/uc~\cite{Ding17p14956}. Our calculated value for the same pathway is 40 meV/uc (Fig.~\ref{Fig3-theory}\textbf{a}), due to the usage of more intermediate images and tighter convergence threshold.
However, a critical issue with NEB is that this method  requires the manual construction of an initial pathway, which can potentially be subject to human bias, and there is no guarantee that such a manually designed pathway can be physically activated by \EOP.
Indeed, we find that this low-barrier pathway involves concerted lateral shifts of entire layers of In and Se atoms (see inset of Fig.~\ref{Fig3-theory}\textbf{a}), coupled with out-of-plane displacements that reverse \POP. We suggest that it is highly improbable for \EOP~to activate this process: In and Se atoms,bearing opposite charges,  should move in opposite directions under the influence of \EOP, rather than moving in the same direction as assumed in NEB calculations. We further perform finite-field NEB calculations with varying \EOP~magnitudes and confirm that $\Delta U^{\rm NEB}$ is weakly dependent on field strength: even an intense \EOP~of 3.0 V/nm (30 MV/cm) only moderately reduces the barrier (Fig.~\ref{Fig3-theory}\textbf{a}). This corroborates with the splitting restriction principle that suggests the absence of a symmetry-adapted polar mode for efficient \POP-\EOP~coupling.

Another important aspect is that $\Delta U^{\rm NEB}$ corresponds to the barrier for a homogeneous switching mechanism during which all dipoles response synchronously to \EOP~without nucleating oppositely polarized domains.
This overlooks the energy costs involved in creating new interfaces that separate the newly formed nucleus and the surrounding domain. 
The difficulty of reversing \POP~in an ideal 2D domain of \ainse~via \EOP~is further corroborated by our large-scale MD simulations that employs a deep neural network-based force field trained with a large database of {\em ab initio} energies and atomic forces from $\approx$25,000 configurations of In$_2$Se$_3$~\cite{Wu21p174107} (see Methods).
We observe that a single-domain monolayer, constructed using an 18000-atom supercell, is immune to \EOP~at 300~K (no switching detected within a simulation period of 3 nanoseconds), even when exposed to giant field strengths achievable in experimental setups. Only at an elevated temperature of 373~K is a nucleus formed with \POP~aligned with \EOP~of 3.0 V/nm (Figs.~S13).
The change in the energy associated with the formation of a 2D nucleus containing $N$ unit cells can be approximated as~\cite{Liu16p360}
\begin{equation}
\Delta U_{\rm nuc}=-2N \mathcal{E}_{\rm OP}p_u+g\sqrt{N}\sigma_i
\label{deltaU}
\end{equation}
where $p_u$ is the out-of-plane electric dipole moment per unit cell, $\sigma_i$ is the (averaged) interface energy per unit-cell length of the boundary that separates the nucleus and the parent domain, and $g$ is a geometric factor depending on the nucleus shape. Detailed derivations for equation~\ref{deltaU} are provided in Supplementary Sect.~VII. Using $p_u=0.09$~$e$\AA~computed with DFT, $\sigma_i=0.23$~eV and $g=8/\sqrt{3}$ both extracted from MD, Fig.~\ref{Fig3-theory}\textbf{b} displays $\Delta U_{\rm nuc}$ as a function of $N$ for various field strengths, and the maximum of the plot determines the nucleation barrier ($\Delta U_{\rm nuc}^*$) and the size of the critical nucleus ($N^*$).
For example, at $\mathcal{E}_{\rm OP} = 1.0$ V/nm, the nucleation barrier is $\approx1.74\times10^5k_BT$, indicating an extremely low nucleation probability at room temperatures even under a high electric field.
The high 1D interfacial energy $\sigma_i$ and the small magnitude of $p_u$~together make it energetically challenging to switch a 2D domain in monolayer \ainse. 
Under experimental conditions, the presence of defects could reduce the nucleation barrier by reducing $\sigma_i$. 

\subsection{Domain-wall-assisted ferroelectric switching}

\begin{figure}[t]
\centering
\includegraphics[width=6.0 in]{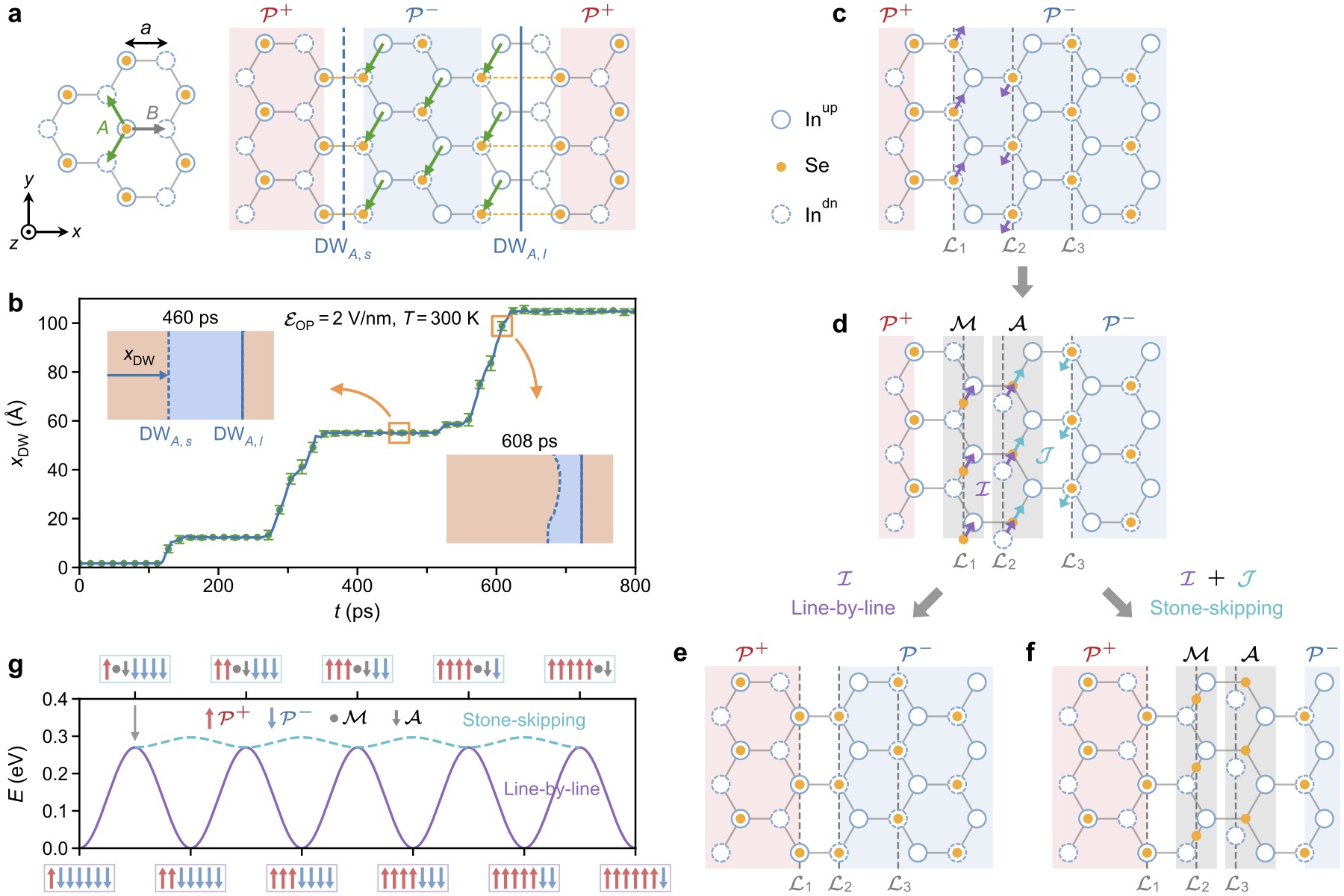}
\caption{{\bf Avalanche dynamics of 1D domain walls driven by \EOP~in 2D $\alpha$-In$_2$Se$_3$.} \textbf{a,} Formation of 1D domain walls. The $180^\circ$ walls separating $\mathcal{P}^+$ and $\mathcal{P}^-$ domains can be created by an in-plane shift of a section of middle-layer Se atoms in either direction $A$ or $B$ by a distance of $a$, accompanied by an out-of-plane shift that flips the polarization. Type-$A$ walls can be categorized as \dwas~characterized by short Se-Se bonds at the wall and \dwal~featuring large Se-Se separations. The green arrows denote the shift directions of Se atoms from their initial positions to the final positions.
\textbf{b,} Temporal evolution of the averaged position ($x_{\rm DW}$) of \dwas~driven by \EOP~obtained from MD simulations of a 4000-atom supercell. \dwal~is immobile while \dwas~moves abruptly and intermittently, indicating avalanche dynamics.
The error bar scales with the geometrical roughness (global width) of the wall. The insets illustrate the flatness of \dwas~at 460 ps and 608 ps (dynamic roughening). 
\textbf{c-f} depict two mechanisms of \dwas~motion derived  from MD. Driven by \EOP, \dwas~in \textbf{c} evolves to a transient interface $\mathcal{MA}$ in \textbf{d} via $\mathcal{P}^-@\mathcal{L}_1\rightarrow\mathcal{M}@\mathcal{L}_1$ and $\mathcal{P}^-@\mathcal{L}_2\rightarrow\mathcal{A}@\mathcal{L}_2$. Immediately following that, two atomic movement patterns denoted as $\mathcal{I}$ and $\mathcal{J}$ in \textbf{d} are possible. The concerted occurrence of $\mathcal{I}$ and $\mathcal{J}$ results in fast sideway movement of $\mathcal{MA}$ in \textbf{f}.
The failure of movement $\mathcal{J}$ recovers the Miller-Weinreich line-by-line mechanism in \textbf{e}. \textbf{g,} Schematics of energy profiles for line-by-line mechanism (solid line) and stone-skipping-like mechanism (dashed line).  The bursts of domain wall motions in \textbf{b} are attributed to the
fast-moving $\mathcal{MA}$ interface and the periods of inactivity are due to the reactivation of \dwas. 
}
\label{Fig2}
\end{figure}

We will now demonstrate that pre-existing 1D domain walls can lower the switching field dramatically.
When viewed from the out-of-plane direction of a single domain, the two layers of In atoms resemble the hexagonal lattice of monolayer $h$-BN with a (projected) bond length of $a$, while middle-layer Se atoms occupy only one type of lattice site (Fig.~\ref{Fig1}\textbf{b}). There exist four variants of 180$^\circ$ domain walls, grouped into two pairs ($A$ and $B$), that can be generated by an in-plane shift of a section of middle-layer Se atoms along the direction connecting In atoms by a distance of $a$ accompanied by an out-of-plane shift that flips the polarization direction. As illustrated in Fig.~\ref{Fig2}\textbf{a}, for each pair, there is a wide domain wall characterized by long Se-Se separations (\eg, \dwal) and a narrow domain wall featuring short Se-Se bonds (\eg, \dwas). We focus on type-$A$ walls for their greater thermodynamic stability compared to type-$B$ walls (see Supplementary Sect.~VIII). Our MD simulations further show that only \dwas~is movable by \EOP; \dwal~is immobile, consistent with its large barrier of 0.7~eV for domain wall motion, as predicted by DFT-based NEB calculations (see Table S2 in Supplementary Material). 
The temporal evolution of the averaged position of \dwas~driven by \EOP~(Fig.~\ref{Fig2}\textbf{b}) unveils two characteristics that differ distinctly from 2D domain walls in bulk ferroelectrics~\cite{Tagantsev10Book,Liu16p360}. First, the motion of \dwas~exhibits avalanche dynamics~\cite{Casals21p345}, wherein this 1D interface moves abruptly and intermittently. The distributions of the lifetimes of avalanche processes can be found in {Supplementary Sect.~IX}.
Second, \dwas~acquires dynamic roughening:
the initially flat wall becomes curved during its motion. The geometrical roughness, represented as error bars in Fig.~\ref{Fig2}\textbf{b}, is characterized by a global width of $\approx$1.8~\AA. 

By analyzing MD trajectories with a fine time resolution down to femtoseconds, we identify a ``stone skipping"-like mechanism that explains the two intrinsic features of \dwas~motion emerged in the absence of defects (disorder pinning). 
As depicted in Fig.~\ref{Fig2}\textbf{c}, in the presence of \EOP, a line of Se atoms ($\mathcal{L}_1$) closest to the boundary move towards the bottom of nearby In$^{\rm up}$ atoms (denoted as $\mathcal{P}^-@\mathcal{L}_1\rightarrow\mathcal{M}@\mathcal{L}_1$).
Simultaneously, a line of In$^{\rm dn}$ atoms ($\mathcal{L}_2$) shift towards those Se atoms of $\mathcal{L}_1$ (denoted as $\mathcal{P}^-@\mathcal{L}_2\rightarrow\mathcal{A}@\mathcal{L}_2$). 
This contrasts with the layer-by-layer switching mechanism pioneered by Miller and Weinreich for sideway motion of 2D domain walls~\cite{Miller60p1460}. Here, the movement of 1D \dwas~engages two lines of atoms and is somewhat reminiscent of the hypothesized ``nuclei stacking" that is assumed to occur in a high-field-induced mobile rough wall (though rarely observed)~\cite{Hayashi72p616}. Importantly, the transient interface, $\mathcal{M}@\mathcal{L}_1$ and $\mathcal{A}@\mathcal{L}_2$, can jointly serve as a moving front (Fig.~\ref{Fig2}\textbf{d}). Specifically, subsequent processes of $\mathcal{A}@\mathcal{L}_2\rightarrow\mathcal{M}@\mathcal{L}_2$ and $\mathcal{P}^-@\mathcal{L}_3\rightarrow\mathcal{A}@\mathcal{L}_3$, achieved via atomic movement pattern 
$\mathcal{J}$
(Fig.~\ref{Fig2}\textbf{d}), effectively move the $\mathcal{MA}$ interface sideways by a distance of $a$ (Fig.~\ref{Fig2}\textbf{f}). At steps where movement $\mathcal{J}$ does not take place, the line-by-line switching mechanism is recovered as the net outcome is $\mathcal{P}^-@\mathcal{L}_1\rightarrow\mathcal{P}^+@\mathcal{L}_1$ (Fig.~\ref{Fig2}\textbf{e}). We propose that the $\mathcal{MA}$ interface is an ``emergent" high-energy interface that only needs to overcome a small kinetic barrier to move (see the dashed line in Fig.~\ref{Fig2}\textbf{g}), analogous to a swiftly skipping stone on water. The probabilistic failure of movement $\mathcal{J}$ could trigger the relaxation of $\mathcal{MA}$ to a flat, low-energy interface (\dwas), the movement of which must overcome a large barrier (see the solid line in Fig.~\ref{Fig2}\textbf{g}). This stone-skipping-like mechanism intuitively explains the bursts of domain wall motions (attributable to the emergent fast-moving $\mathcal{MA}$ interface) separated by periods of inactivity (due to the reactivation of \dwas). The dynamic roughening of \dwas, manifested as a wavy moving font (see MD snapshots in Fig.~S17), is a consequence of the interplay between two distinct mechanisms of domain wall motion along an extended 1D interface. Across larger length scales, specific sections of the interface progress through the slower, line-by-line mechanism, while other segments advance more rapidly via the stone-skipping-like mechanism. This differential speed of movement along the interface's length results in the curvature during its motion. We note that during the period when
the average position of the wall remains unchanged, the wall still exhibits significant dynamics: the
interface fluctuates around the average position (see Fig.~S19). This behavior is indicative of a
thermally-driven activation process.

\begin{figure}[t]
\centering
\includegraphics[width=5 in]{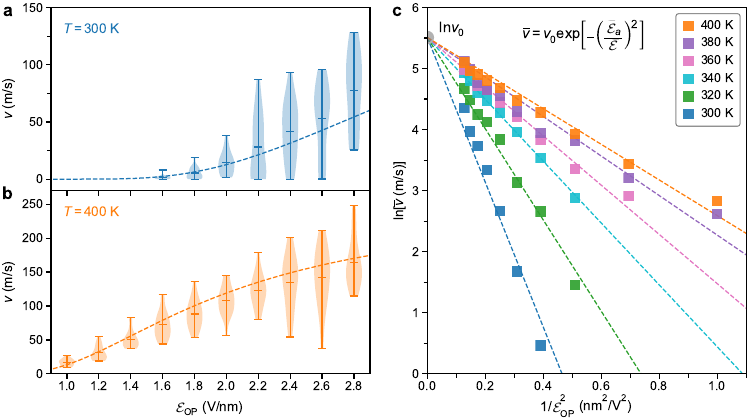}
\caption{{\bf Velocity of 1D domain walls driven by \EOP~in 2D $\alpha$-In$_2$Se$_3$.} 
Violin plot of domain wall velocity ($v$) at \textbf{a,} 300~K and \textbf{b,} 400~K. The mean velocity $\bar{v}$ is fitted to equation (2) that describes a creep process. 
We find $v_0=247$ m/s and $\mu=2$. \textbf{c,} Plot of ln($\bar{v}$) versus $1/{\mathcal{E}_{\rm OP}^2}$ curves for different temperatures. All linear fits (dashed lines) converge at the same intercept corresponding to $v_0=247$ m/s.}
\label{v_dw}
\end{figure}

We use MD simulations to quantitatively estimate the velocity ($v$) of \dwas~over a wide range of temperatures ($T$) and \EOP. 
It is noted that the value of $v$ is calculated by measuring the distance that the wall traverses within a specific time frame, while also taking into account the periods of inactivity when the wall is stationary. Due to the inherent avalanche dynamics, the values of $v$, obtained from multiple MD trajectories, exhibit a considerable fluctuation for a specific \EOP~and $T$, as shown by the violin plots at 300 K and 400 K (Fig.~\ref{v_dw}\textbf{a}-\textbf{b}). Remarkably, even in the absence of defects, we discover that the mean 1D domain wall velocity, $\bar{v}$, can be well described with a creep process as ~\cite{Tybell02p097601, Jo09p045701}
\begin{equation}
\bar{v}=v_0\exp\left[-\left(\frac{\mathcal{E}_a}{\mathcal{E}}\right)^{\mu}\right],
\end{equation}
where $v_0$ is the domain wall velocity under an infinite field, $\mathcal{E}_a$ is the temperature-dependent activation field. Within a statistical description of domain wall motion as a critical phenomenon, that is, an elastic interface moving in random media, $\mu$ is the creep
exponent that depends on the dimensionality of the interface and the universality class of
the disorder landscape pinning the interface~\cite{Lemerle98p849}. As depicted in Fig.~\ref{v_dw}\textbf{c}, the plots of $\ln(v)$ versus $1/\mathcal{E}_{\rm OP}^2$~curves yield linear relationships  across a broad spectrum of field strengths (1.0--2.8 V/nm) and various temperatures (300--400~K). Notably, all linear fits converge at the same intercept that corresponds to $v_0= $ 247 m/s. This serves as a strong evidence for an unusual creep exponent, $\mu=2$, higher than the well-known value of $\mu=0.25$ for 1D magnetic domain walls in ultrathin magnetic films~\cite{Lemerle98p849} as well as $\mu=1$ for 2D ferroelectric domain walls in typical perovskite ferroelectrics~\cite{Shin07p881, Liu16p360}. We propose \EOP-driven creep motion of \dwas~in monolayer \ainse~probably belongs to an entirely new universality class. A higher value of creep exponent implies that this interface is more sensitive to changes in the magnitude of \EOP, especially at low driving forces, which is advantageous for fine-tuning domain wall mobility. Moreover, we observe
the absence of an intrinsic creep–depinning transition, even at a colossal field strength of 2.8 V/nm (28 MV/cm), likely attributable to the small Born effect charges (0.49 for In and $-0.45$ for Se) in the out-of-plane direction: the weak \EOP-\POP~coupling is insufficient to reduce the barrier of domain wall motion compared to thermal fluctuations, thus leading to persistent creep behavior of \dwas.

\subsection{Domain wall motion driven by in-plane electric fields}

The movement of \dwas~driven by \EOP~involves in-plane displacements of In and Se atoms near the boundary (Fig.~\ref{Fig2}\textbf{c-f}), hinting at a potential coupling between \dwas~and in-plane fields. Considering the symmetry, the presence of \dwas~locally disrupts the C$_{\rm 3v}$ point group symmetry, in turn generating local dipoles in \dwas~that could be coupled to \EIP. This is confirmed by our finite-field MD simulations, which demonstrate \EIP-driven \dwas~motion. Importantly, owing to the large in-plane Born effective charges of Se and In atoms, applying \EIP~of 0.12 V/nm is sufficient to drive the movement of \dwas~within 100 ps at 300 K. This field strength is notably lower than what would be needed when \EOP~is applied, and is comparable to the field strengths ($\approx$0.2 V/nm) used in experiments~\cite{Wang20p45} . 

\begin{figure}[t]
\centering
\includegraphics[width=6.4 in]{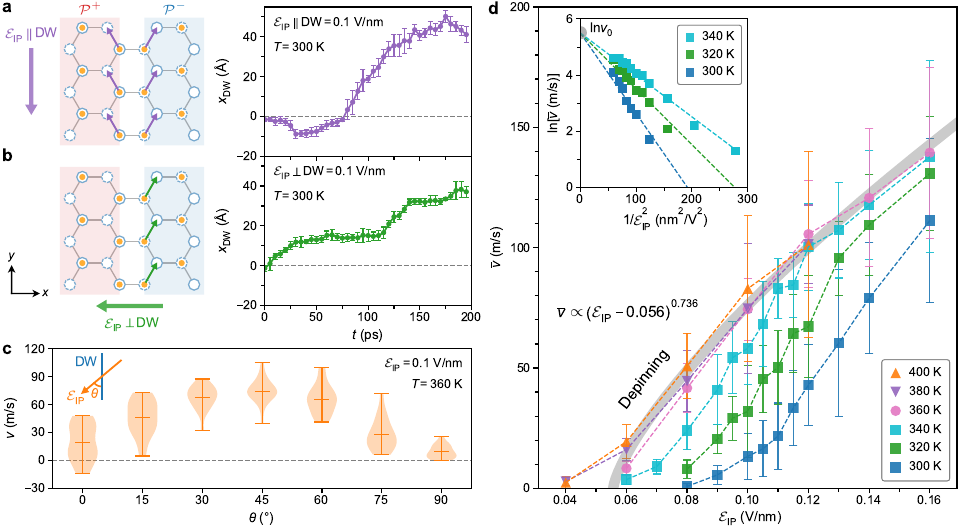}
\caption{{\bf Dynamics of 1D domain walls driven by \EIP~in 2D $\alpha$-In$_2$Se$_3$.} 
Temporal progression of \dwas~in the presence of \EIP~applied \textbf{a,} parallel to the wall ($\mathcal{E}_{\rm IP}||{\rm DW}$) and \textbf{b,} perpendicular to the wall ($\mathcal{E}_{\rm IP}\perp {\rm DW}$), revealing avalanche dynamics in both cases.
The field strength is 0.1~V/nm.  \textbf{c,} Violin plot of domain wall velocity as a function of the relative orientation ($\theta$) between \EIP~and the wall at 360~K. \textbf{d,} $T$- and \EIP-dependent mean velocity of \dwas~for $\theta=45^\circ$.  The velocity data at temperatures above 340 K and $\mathcal{E}_{\rm IP}\ge0.06$ V/nm demonstrate depinning characteristics and is fitted to equation (3). We find ${\mathcal{E}_{\rm IP}^{\rm C0}}=0.056$~V/nm and $\Theta=0.736$ (bold gray line).
Velocity data within the lower temperature range (300--340~K) and lower field range (0.06--0.12~V/nm) show creep behavior. The linear relationship (dashed lines) between ln($\bar{v}$) and $1/{\mathcal{E}_{\rm IP}^2}$, as shown in the inset, confirms a creep exponent of $\mu=2$. The linear fits also converge at the same intercept that corresponds to $v_0=247$ m/s.
}
\label{Fig4}
\end{figure}

The \EIP-driven \dwas~motion is found to exhibit a pronounced field-orientation dependence. When \EIP~is applied parallel to \dwas~($\mathcal{E}_{\rm IP}||{\rm DW}$), this 1D interface shows an equal likelihood of moving either to the left or the right, corresponding microscopically to the equally probable hopping processes of Se atoms situated in proximity to \dwas~(Fig.~\ref{Fig4}\textbf{a}). 
This behavior is a manifestation of spontaneous symmetry breaking: while \EIP~does not inherently favor any particular direction due to the lack of physical in-plane polarization, the wall opts for a specific direction, which is induced by thermal activation. The temporal progression of the domain wall position (Fig.~\ref{Fig4}\textbf{a}) clearly demonstrates that \dwas~retains the avalanche dynamics and oscillates back and forth in a stochastic manner in the case of $\mathcal{E}_{\rm IP}||{\rm DW}$. In comparison, when \EIP~is applied perpendicular to \dwas~($E_{\rm IP}\perp {\rm DW}$), the wall moves deterministically, in the direction counter to that of the Se atom hopping (Fig.~\ref{Fig4}\textbf{b}). We map out $\bar{v}$ as a function of the relative orientation ($\theta$) between \EIP~(of 0.1~V/nm) and the wall at 360~K, showing a strong anisotropy with the maximum velocity peaking at $\theta \approx 45^\circ$ (Fig.~\ref{Fig4}\textbf{c}). Such highly tunable domain wall mobility is beneficial for controlled design of polarization switching speed.

The $T$- and \EIP-dependent mean velocity of \dwas~for $\theta$=45$^\circ$ is displayed in Fig.~\ref{Fig4}\textbf{d}, indicating an intrinsic creep-depinning transition~\cite{Liu16p360}. In particular, the velocity data at temperatures exceeding 340~K overlap when $\mathcal{E}_{\rm IP}>0.06$~V/nm, a hallmark feature of depinning wherein the velocity becomes temperature independent.
These velocity data can be fitted to 
\begin{equation}
    \bar{v} \propto (\mathcal{E}_{\rm IP}-\mathcal{E}_{\rm IP}^{\rm C0})^\Theta,
\end{equation}
where $\Theta$ is a velocity exponent and $\mathcal{E}_{\rm IP}^{\rm C0}$ is the crossing field at zero Kelvin. The fitting yields $\Theta=0.736$, a value nearly identical to the reported velocity exponent of 0.72 for 90$^\circ$ domain walls in PbTiO$_3$. This implies the existence of universal depinning dynamics for domain walls in 2D and 3D ferroelectrics. At temperatures of 340~K or lower, the velocity has a strong dependence on temperature and exhibits a nonlinear relationship with \EIP, suggesting a creep behavior. As presented in the inset of Fig.~\ref{Fig4}\textbf{d}, the velocity data in the creep region can be well described by equation (2) using the same parameters, $v_0=247$ m/s and $\mu=2$, as derived from \EOP-driven \dwas~motions shown in Fig.~\ref{v_dw}\textbf{c}. The velocity data collected under both \EIP~and \EOP~affirm the robustness of an atypical creep exponent of 2, supporting the presence of a new universality class. 

\section{Discussion}
Our investigation, which integrates both experimental characterizations and theoretical
simulations on the ferroelectric switching in 2D \ainse,addresses the three aforementioned questions. First, PFM studies following rigorous protocols yield conclusive evidence indicating the absence of measurable \PIP~in \ainse. Specifically, the comparative study of angle-resolved IP-PFM on both 2H \ainse~and NbOI$_2$ demonstrates that the measured lateral response in \ainse~does not originate from intrinsic in-plane piezoresponse, but is due to a crosstalk artifact from the out-of-plane signal.

Second, the forbidden splitting of Wyckoff orbits for the transition from paraelectric \binse~to ferroelectric \ainse~introduces a trade-off between the robustness of \POP~and the difficulty in switching it: the lack of a symmetry-adapted polar mode renders \POP~stable against the depolarization field but also makes \EOP-\POP~coupling rather inefficient. This splitting restriction principle, along with the resultant difficulty of \EOP-driven polarization reversal, is expected to be a prevalent characteristic across a wide spectrum of vdW bilayers exhibiting sliding ferroelectricity and mori\'e ferroelectricity. 
The existence of domain walls, particularly \dwas, is found to be essential for the ferroelectric switching driven by \EOP~in 2D \ainse. The giant magnitude of \EOP~(at the order of 1 V/nm) required for moving domain walls, as predicted by our MD simulations and corroborated by experiments~\cite{Xiao18p227601}, 
can be attributed to the weak \EOP-\POP~coupling. 

Despite the lack of switchable \PIP~in a single-domain monolayer, the presence of \dwas~breaks the out-of-plane three-fold rotational symmetry on a local level, allowing for a coupling between local dipoles of \dwas~and \EIP. Benefiting from the large in-plane Born effective charges and the unique mechanism of \dwas~motion involving in-plane atomic displacements, applying \EIP~proves to be more efficient than using \EOP~for inducing ferroelectric switching, resolving a longstanding inconsistency between experimental observations and theoretical predictions. 

Finally, the 1D ferroelectric domain wall exhibits unusual avalanche dynamics when subjected to electric fields, stemming from a competition between two distinct domain wall motion mechanisms: the conventional  Miller-Weinreich line-by-line mechanism and the stone-skipping-like mechanism facilitated by an emergent high-energy interface. Our extensive dataset of domain wall velocity at various temperatures and field strengths (including both \EOP~and \EIP) reveals an unprecedented creep exponent of $\mu=2$, diverging from the known values for magnetic domain walls~\cite{Lemerle98p849} and ferroelectric domain walls in bulk perovskite ferroelectrics~\cite{Shin07p881, Liu16p360}. 

Creep motion of a $d$-dimensional elastic interface moving in a $d+1$-dimensional random media is an important physical behavior presented in a vast range of diverse systems such as vortices in type-II superconductors~\cite{Blatter94p1125}, density waves~\cite{Fukuyama78p535}, burning~\cite{Myllys00p1946} and wetting fonts~\cite{Rubio89p1685}, and cell migration~\cite{Chepizhko16p11408}. The indication of an entirely new universality class associated with $\mu=2$, as reported here, highlights the potential of 1D ferroelectric domain walls in 2D ferroelectrics for in-depth understanding of the fundamental physics of moving interfaces in reduced dimensions. Furthermore, the strong anisotropic response of domain walls to external fields and tunable onset field for the creep-depinning transition can be harnessed to configure the switching speed in 2D \ainse. 
Given the structural similarity between monolayer \ainse~and vdW bilayers with sliding ferroelectricity, we anticipate the atomic-level insights and quantitative understanding of domain wall mobility presented here will contribute to the comprehension of intrinsic ferroelectric switching in a broad range of 2D ferroelectrics, ultimately leading to enhanced functionalities of 2D ferroelectric-based devices.

\section{Methods}

\textbf{DFT calculations}  
\indent All first-principles density functional theory calculations are performed using the Vienna {\em ab initio} Simulation Package (\texttt{VASP})~\cite{Kresse96p11169,Kresse96p15}. The interaction between core ions and electrons is modeled using the projector augmented wave (PAW) method~\cite{Blochi94p17953}. We choose the PBE functional as the exchange-correlation functional~\cite{Perdew08p136406}, and the van der Waals (vdW) interaction is considered using the Grimme method with a zero-damping function~\cite{Grimme10p154104}. A slab model incorporating a vacuum layer with a thickness exceeding 20~\AA~ along the $z$-axis is used to model monolayer In$_2$Se$_3$; the dipole correction is employed to counteract spurious interactions between periodic images across the vacuum layer. DFT calculations are based on an energy cutoff of 700 eV, a 7$\times$7$\times$1 Monkhorst–Pack $k$-point mesh for Brillouin zone sampling of the unit cell, and an energy convergence threshold of 10$^{-8}$ eV for electronic self-consistency. The convergence criterion for structure optimization is set to 10$^{-7}$eV in energy. An external electric field is modeled by introducing a planar dipole layer at the center of the vacuum region~\cite{Neugebauer92p16067}. The climbing image nudged elastic band (CL-NEB) method~\cite{Henkelman00p9901} with force convergence set to 0.01 eV/\AA~is used to determine the minimum energy path for polarization reversal.

\textbf{Force field of 2D $\alpha$-In$_2$Se$_3$}  
\indent The force field utilized in our large-scale molecular dynamics simulations is a deep neural network-based model potential, known as deep potential (DP)~\cite{Zhang18p143001,Zhang18p4441}. The DP model maps an atom's local environment to its energy ($E_i$), where the total energy $E$ is obtained as $E=\sum_iE_i$. 
We have successfully trained a DP model of monolayer In$_2$Se$_3$ using a concurrent learning process that effectively updates the first-principles-based training database~\cite{Wu21p174107}. 
The neural network architecture consists of three hidden layers, with each layer containing 240 neurons. The final training database comprises 22,600 monolayer configurations and 2,163 bulk structures. 
With $10^6$ learnable parameters, the model has demonstrated
remarkable accuracy in fitting to DFT energies and atomic forces (Fig.~S8).
For the atomic forces of all atoms in the entire training database, 
the mean absolute error (MAE) and root mean squared error (RMSE) are 
0.068 and 0.096 eV/{\AA}, respectively. 
The model potential is capable of predicting a variety of thermodynamic properties of In$_2$Se$_3$ polymorphs (including $\alpha$, $\beta$, and $\beta'$), the DFT potential energy surface for the in-plane sliding of the central Se sublayer, temperature-driven phase transitions, and energy profiles for polarization reversal and 180$^\circ$ domain wall motions in monolayer $\alpha$-In$_2$Se$_3$ (Fig.~S9). We note that all local
atomic environments involved in domain wall motions are well-represented in the training database (Fig.~S10).

\textbf{Molecular dynamics} 
\indent We perform constant-temperature constant-volume ($NVT$) deep potential molecular dynamics simulations using LAMMPS~\cite{Plimpton95p1}. The temperature is maintained using the Nos\'e-Hoover thermostat and the time step is set to 1 femtosecond. The free-standing monolayer $\alpha$-In$_2$Se$_3$ containing type-$A$ domain walls is modeled using a $20\times40\times1$ supercell comprised of 4,000 atoms and a vacuum layer of 22~\AA~under periodic boundary conditions, as shown in Fig.~S16. The effects of electric fields are modeled in classical MD simulations through the force method~\cite{Umari02p157602}. This approach involves adding an additional force to each ion, calculated as the product of the ion's Born effective charge and the magnitude of the electric field.
We note that at finite temperatures, the unstained free-standing monolayer $\alpha$-In$_2$Se$_3$ can spontaneously transition into $\beta'$-In$_2$Se$_3$ due to their close thermodynamic stability, especially in the presence of domain walls and/or external in-plane electric fields. This is actually consistent with several experimental observations that the phase transition from $\alpha$-In$_2$Se$_3$ to $\beta'$-In$_2$Se$_3$ can be induced by strain~\cite{Zheng22peabo0773,Han22p55}.
To explore the intrinsic switching dynamics of 1D domain walls in monolayer $\alpha$-In$_2$Se$_3$, we imposed a minor tensile strain ($\approx$1.6\%) to stabilize the {$\alpha$} phase and the domain walls. At a specific temperature, an equilibrium run lasting at least 1 nanosecond (ns) is executed before the electric field is introduced to initiate domain wall movement and to measure the velocity ($v$) of the domain wall. The value of $v$ is determined directly by tracking the distance traveled by the wall within a specific time frame, while also taking into account the periods of inactivity when the wall is stationary to capture the effect of avalanche dynamics.
We carry out 20 independent runs (ranging from 0.1 to 1 ns) for specific temperature and electric field strength to obtain the velocity distribution of the domain wall. The methodology for determining the domain wall position during its motion relies on the coordination numbers of Se atoms. At a specific time point $t$ (in picosecond), we first determine the averaged structure by averaging configurations sampled every 0.1~ps within a time interval from $t-\delta$ to $t$, where $\delta=$1--5~ps depending on the domain wall speed. This treatment reduces the thermal noise. We then search for the Se-In coordination within a radius of 4.2 \AA~centered around the Se atom. If the coordination number of a Se atom exceeds 4, we record its position, and these Se atoms define the position of the domain wall. In an effort to improve reproducibility, we have made our training database, force field model, training metadata, essential input and output files publicly available in an open repository~\cite{notebook-In2Se3}.

\textbf{Piezoresponse force microscopy}
\indent The PFM studies are conducted using a commercial atomic force microscope (Oxford Instrument, MFP-3D origin+). Dual AC resonance tracking (DART) method is employed for PFM imaging and local switching spectroscopy measurements. Two types of conductive AFM probes with spring constants of $\approx 2.8$ N/m (Asylum Research, ASYELEC.01-R2) and $\approx 0.2$ N/m (Nanoworld, CONTPt) are used for cross validation and comparison. Micro-Raman spectroscopy and mapping are measured by an integrated confocal Raman microscope (Horiba, Xplora Plus) using a 532 nm laser focused by 50X long working distance objective. 


\begin{acknowledgments}
L.B., C.K., T.Z., and S.L. acknowledge the supports from National Natural Science Foundation of China (52002335) and Westlake Education Foundation. The computational resource is provided by Westlake HPC Center. We thank Dr.~Jiawei Huang for useful discussions regarding the categorization of domain walls.
\end{acknowledgments}

\clearpage
\newpage
\bibliography{SL_updated.bib}

\clearpage
\newpage

\clearpage
\newpage

\clearpage
\newpage

\clearpage
\newpage

\clearpage
\newpage

\clearpage
\newpage


\end{document}